# Photon Acceleration in a Flying Focus

A.J. Howard, D. Turnbull, A.S. Davies, P. Franke, D.H. Froula, and J.P. Palastro
*University of Rochester, Laboratory for Laser Energetics, Rochester NY, USA*

**Abstract**

A high-intensity laser pulse propagating through a medium triggers an ionization front that can accelerate and frequency-upshift the photons of a second pulse. The maximum upshift is ultimately limited by the accelerated photons outpacing the ionization front or the ionizing pulse refracting from the plasma. Here we apply the flying focus—a moving focal point resulting from a chirped laser pulse focused by a chromatic lens—to overcome these limitations. Theory and simulations demonstrate that the ionization front produced by a flying focus can frequency-upshift an ultrashort optical pulse to the extreme ultraviolet over a centimeter of propagation. An analytic model of the upshift predicts that this scheme could be scaled to a novel table-top source of spatially coherent x-rays.

A growing number of scientific fields rely critically on high intensity, high-repetition rate sources of extreme ultraviolet (XUV) radiation (wavelengths < 120 nm). These sources provide high-resolution imaging for high energy density physics and nanotechnology [1,2], fine-scale material ablation for nanomachining, spectrometry, and photolithography [3-5], and ultrafast pump/probe techniques for fundamental studies in atomic and molecular physics [6-8]. While XUV sources have historically been challenging to produce, methods including nonlinear frequency mixing [9], high harmonic generation [10,11], and XUV lasing or line emission in metal-vapor and noble-gas plasmas [5,12] have demonstrated promising results. Despite their successes, each of these methods introduces tradeoffs in terms of tunability, spatial coherence, divergence, or efficiency [5,9-12]. Photon acceleration offers an alternative method for tunable XUV production that could lessen or even eliminate these tradeoffs.

Photon acceleration refers to the frequency upshift of light in response to a refractive index that decreases in time [13,14]. In analogy to charged particle acceleration, the increase in photon energy, i.e. frequency, accompanies an increase in group velocity. In the context of an electromagnetic pulse, the leading phase fronts experience a higher index than adjacent, trailing phase fronts, which manifests as a local phase velocity that increases over the duration of the pulse. The trailing phase front, on account of its higher phase velocity, gradually catches up with the leading front, compressing the wave period. In a medium with normal dispersion, the resulting frequency upshift translates to an increase in the local group velocity.

Plasmas, in particular, provide an ideal medium for photon acceleration: the refractive index depends on the density of free electrons, which can be rapidly increased or decreased over time through ionization and recombination or manipulated through electrostatic wave excitation. Specifically, a photon of frequency $\omega$ in an isotropic plasma experiences a refractive index

$$n(\omega) = \left(1 - \frac{\omega_p^2}{\omega^2}\right)^{1/2}, \quad (1)$$

where $\omega_p = (e^2 n_e / m\varepsilon_0)^{1/2}$ is the plasma frequency, $n_e$ the free electron density, $e$ the electron charge, $m$ the electron mass, and $\varepsilon_0$ the permittivity of free space. An increase in the electron density over time—for example by ionization—provides a decreasing refractive index that will accelerate the photons of a co-located pulse.

A prototypical scheme for photon acceleration involves propagating a witness pulse in an ionization front triggered by a copropagating drive pulse [15-20]. In spite of the impressive frequency shifts (>10×) predicted by theory and simulations [13-16,21-23], experiments in the optical regime have met with limited success (~1.25×) [17,18,20,24] on account of witness pulse refraction and drive pulse diffraction. While these effects can be remedied by preforming a plasma or pre-shaping a gas to provide a guiding structure [25-27], two inherent limitations to the upshift remain. First, the drive pulse, and hence the ionization front, travels at a subluminal group velocity. As the witness pulse accelerates, it quickly outpaces the ionization front, terminating the interaction. Second, the drive pulse refracts from the plasma it creates, i.e. it undergoes ionization refraction, limiting the formation of a continuous ionization front.

In this Letter, we demonstrate, for the first time, a scheme for photon acceleration within a copropagating ionization front that shifts an optical pulse to the XUV. The scheme utilizes a novel photonic technique known as the *flying focus* to overcome the aforementioned limitations [28,29]. An appropriately chirped drive pulse, focused through a chromatic lens, exhibits an intensity peak that counterpropagates at the speed of light in vacuum, $c$, with respect to its group velocity [28,30,31]. The peak intensity, in turn, triggers an ionization front travelling at $c$, which can continually accelerate the photons of a copropagating witness pulse. A schematic is displayed in

Fig. 1 for the case of a diffractive optic. The peak intensity of the drive pulse has a self-similar profile as it travels through the focal region, $z_f$, at the focal velocity, $v_f$:

$$z_f = \frac{\Delta \lambda}{\lambda_c} f, \quad (2)$$

$$v_f = \left(1 + \frac{v_d T}{z_f}\right)^{-1} v_d, \quad (3)$$

where $\lambda_c$ is the central wavelength of the drive pulse, $\Delta \lambda / \lambda_c$ its fractional bandwidth, $f$ the focal length of the diffractive optic at $\lambda_c$, $v_d = cn$ the group velocity, and $T$ the stretched pulse duration.

By decoupling the ionization-front velocity from the group velocity of the drive pulse, this scheme removes both of the inherent limitations of the prototypical photon accelerator. Most notably, the interaction distance is no longer limited by outpacing, as the accelerated photons can never outpace a luminal (traveling at $c$) ionization front. The interaction distance is solely determined by Eq. (2) and can be extended long past the Rayleigh range of any single frequency component within the drive pulse. Second, counter-propagating the drive pulse with respect to the ionization front mitigates ionization refraction, as the focus of the drive pulse only encounters the un-ionized medium [30,31].

Figure 2 demonstrates that photon acceleration in the ionization front formed by the flying focus can upshift the frequency of an 87 fs witness pulse from the optical ($\lambda = 400$ nm) to the XUV ($\lambda = 91$ nm). The figure shows four snapshots from a photon kinetics simulation. Each snapshot is shown in the moving frame $\xi = ct - z$, where $t$ is the time elapsed after injecting the witness pulse. The photons of the witness pulse enter the ionization front in Fig. 2a, each with an initial vacuum wavelength of 400 nm. The photons continually upshift in frequency as they copropagate with a temporal gradient in electron density as seen in Figs. 2b and c. After ~1 cm, the photons—now upshifted to a minimum vacuum wavelength of 91 nm—approach the end of the focal region and encounter a decelerating ionization front, terminating their upshift.

Figure 3 displays the evolution of the vacuum wavelength throughout the focal region and the final dispersion of the pulse. The witness pulse acquires a broad bandwidth of 83 nm, but retains an ultrashort duration, ~150 fs. The vacuum wavelengths within the output pulse increase nearly quadratically from 91 nm, at the leading edge, to 174 nm at the trailing edge. The dispersion results from the unique inhomogeneity in the temporal gradient experienced by each photon along its path, as seen in Fig. 2. The nonlinear inhomogeneity results in substantially less temporal stretching than an idealized linear gradient: ~150 fs compared to ~380 fs [15].

Initially the vacuum wavelength rapidly decreases. The upshifting, however, gradually slows due to the weakening plasma response: $n^2 - 1 = -\omega_p^2 / \omega^2$. An analytic model, shown as the black dashed curve in Fig. 3a, predicts this effect. Assuming an electron density profile with a constant gradient moving at $c$, the vacuum wavelength evolves according to

$$\lambda(z) = \left(1 + \frac{\omega_{p0}^2}{\omega_0^2} \frac{z}{L_T}\right)^{-1/2} \lambda_0 \quad (4)$$

where $\lambda_0$ is the initial wavelength of a photon in the witness pulse, $\omega_0 = 2\pi c / \lambda_0$, and $\omega_{p0}$ is the value of the plasma frequency at $\xi = L_T$. The value of $\omega_{p0}^2 / L_T$ used in Fig. 3a was averaged over the path of the highest-frequency photon. The analytic curve is in good agreement with the simulation. Discrepancies result from photons initiated with sub-optimal delays with respect to the drive pulse and small variations in the ionization front velocity due to the self-consistent propagation of the drive pulse.

Inspection of Eq. (4) reveals several paths to shorter wavelengths: increasing the interaction length, increasing the peak electron density, or decreasing the scale length. The interaction length [Eq. (2)] can be extended by increasing the bandwidth of the drive pulse or the focal length; the peak electron density can be increased by propagating within a higher density media, such as solid-density targets (~$10^{22}$ cm$^{-3}$); the scale length can be decreased by increasing the intensity of the drive pulse or decreasing the effective duration of the flying focus intensity peak such that ionization occurs more rapidly.

Figure 4 illustrates the efficacy of this scheme compared to the prototypical photon accelerator and its sensitivity to the focal velocity. Without a flying focus, the drive pulse co-propagates with the witness pulse. To emulate this case, two generous simplifications are made in the simulation: First, ionization refraction of the drive pulse is ignored. Second, the ionization front travels at a group velocity $v_d = 0.96c$—the group velocity corresponding to only one third of the peak electron density. Even with these advantages, the prototypical photon accelerator only achieves a minimum vacuum wavelength of 245 nm (black solid curve), a fractional shift of ~1.6 compared to ~4.4 when using a flying-focus. Additionally, when $v_f = c$, the wavelength could be further downshifted by extending the focal region, $z_f$, whereas in the prototypical accelerator, no further frequency conversion is possible.

Table 1 lists the parameters for the drive pulse used to create the luminal ionization front. The parameters correspond to a frequency-doubled Ti:sapphire laser propagating in hydrogen gas. A conservative value of bandwidth was chosen to ensure an ionization front with a constant focal velocity, which requires a relatively flat band within the spectral energy density. The focal length was chosen such that, when using 6 nm of bandwidth, the longitudinal focal region [Eq. (2)] would be large enough ($z_f > 1$ cm) to ensure significant wavelength shifts, but small enough to remain computationally viable. After fixing the length of the focal region, the focal velocity was tuned by adjusting the chirp of the drive pulse. The blue dashed-dot and red dashed curves in Fig. 4 illustrate the consequences of tuning the chirp for too low or too high of a focal velocity, respectively. When too low, the witness pulse quickly outruns the ionization front; when too high, the ionization front outruns the witness pulse. In either case, the wavelength shift is reduced. Notably, the subluminal focal velocity still outperforms the prototypical accelerator.

The electron density profile used in the photon kinetics simulations was extracted from self-consistent simulations that capture the evolution of the drive pulse and the ionization dynamics of the medium in which it propagates [30]. Ultimately, the electron density profile is generated and modified by field ionization, collisional ionization, radiative recombination, and three-body recombination, while the temperature evolves through inverse Bremsstrahlung absorption and ionization cooling.

A high-density gas was used to maximize the temporal gradient of the electron density. At a gas density of $1.75 \times 10^{21}$ cm$^{-3}$, it was found that 5.6 J of pulse energy yielded a near maximum electron density (~$1.4 \times 10^{21}$ cm$^{-3}$), with further increases in energy providing diminishing returns. While more robust to ionization refraction than a traditional pulse, at such high densities the flying focus began to undergo plasma refraction, preventing the peak intensity from (1) reaching the plasma and (2) providing a heat source to overcome cooling from impact ionization. With the high density and cooling, three body recombination limited the maximum density. Nevertheless, the flying focus pulse created a sharp gradient over the entire focal region, ~1 cm, a distance nearly two orders of magnitude greater than the Rayleigh range (~366 μm).

With the density profile from the propagation simulations, photon kinetics equations were used to determine the trajectories of photons within the witness pulse. The photon dispersion relation, $\omega = (\omega_p^2 + c^2 k_z^2)^{1/2}$ where $k_z$ is the wavenumber parallel to the propagation axis, results from the lowest order Eikonal approximation to the wave equation for the electric field of the witness pulse. Treating the dispersion relation as the photon Hamiltonian provides equations of motion for spatial refraction, the group velocity, and frequency conversion:

$$\frac{dk_z}{dz} = \frac{1}{2c^2 k_z}\left(\frac{\partial}{\partial \xi} - \frac{\partial}{\partial z}\right)\omega_p^2, \quad (5)$$

$$\frac{d\xi}{dz} = \frac{\omega}{ck_z} - 1, \quad (6)$$

$$\frac{d\omega}{dz} = \frac{1}{2ck_z}\frac{\partial \omega_p^2}{\partial \xi}. \quad (7)$$

At next order, the Eikonal approximation yields the transport equation for the photon energy density:

$$\frac{\partial E_0^2}{\partial t} + \nabla \cdot (v_g E_0^2) = -\frac{1}{\omega}\frac{d\omega}{dt} E_0^2 \quad (8)$$

where $E_0$ is the electric field amplitude of the witness pulse. According to Eq. (8), the quantity $G = \omega E_0^2$ is conserved along the photon trajectory: as the witness pulse upshifts in frequency, it necessarily decreases in energy density, translating to a loss in photon number. Physically, the witness pulse losses energy by imparting the electrons within the ionization front with a drift momentum as they are born.

Using Eq. (8), the energy efficiency of the photon acceleration process demonstrated in Fig. 3a is approximately 25%, as the frequency is shifted by a factor of ~4. However, this 25% efficiency estimate applies only to the witness pulse itself. A far more conservative estimate of energy efficiency results when considering both the witness pulse *and* the drive pulse. To estimate this overall efficiency, consider a witness pulse with an intensity just below the photoionization threshold—an intensity low enough to ensure that the drive pulse accounts for nearly all of the photoionization and heating that sustains collisional ionization. The propagation simulations show that a drive pulse described by the parameters in Table I creates an approximately 50 μm diameter ionization front. Using a square-pulse approximation, the witness pulse energy is estimated as $E_w = I_w A_w T_w$ where $I_w$ is the witness pulse peak intensity (~ $1.0 \times 10^{14}$ W/cm$^2$), $A_w$ the area of the ionization front (~ 2000 μm$^2$), and $T_w$ the witness pulse duration (~ 87 fs), providing $E_w$ ~ 170 μJ. For a total input energy of 5.6 J and an output of 43 μJ in the XUV, the overall efficiency becomes roughly 10$^{-5}$. With an output of 43 μJ, this scheme compares favorably to plasma-based soft x-ray lasers and previous experiments of photon acceleration demonstrating more modest wavelength shifts (from $\lambda$ = 620 nm to 604 nm) [8,20].

A novel scheme for photon acceleration in a luminal ionization front has been shown to upshift optical photons to the XUV. By applying the recently demonstrated spatiotemporal technique, the "flying focus," the scheme eliminated two inherent limitations of a prototypical photon accelerator: the outpacing of the front by the accelerated photons and ionization refraction of the drive pulse. Photon kinetics simulations for an 87 fs witness pulse copropagating with an ionization front traveling at $c$ demonstrate frequency conversion from the optical ($\lambda = 400$ nm) to the XUV ($\lambda = 91$ nm). Shorter vacuum wavelengths are attainable by increasing the bandwidth of the drive pulse (> 6 nm), the focal length (> 1 m), the intensity of the drive pulse (> $10^{17}$ W/cm$^2$), or moving to solid-density targets (> $10^{21}$ cm$^{-3}$). As a result, this scheme represents a promising method for the production of spatially coherent x-rays at the table-top scale.

## Acknowledgements

The authors would like to thank W. Knox, K. Qu, M. R. Edwards, and N. J. Fisch for constructive discussion. This material is based upon the work supported by the U.S. Department of Energy Office of Fusion Energy Sciences under Contract Nos. DE-SC0016253 and DE-SC0019135, the Department of Energy National Nuclear Security Administration under Award No. DE-NA0001944, the University of Rochester, and the New York State Energy Research and Development Authority.

| Pulse Parameters | Value | Medium Parameters | Value |
|---|---|---|---|
| central wavelength | 400 nm | species | $H_2$ |
| bandwidth | 6 nm | density | $1.75 \times 10^{21}$ cm$^{-3}$ |
| duration (2$^{nd}$ moment) | 87 fs | ionization energy | 13.6 eV |
| stretched duration | 54 ps | | |
| energy | 5.6 J | | |
| radial pulse shape | SG 8 | | |
| spectral pulse shape | SG 10 | | |
| focal length | 1.02 m | | |
| f-number | 14 | | |

Table 1. Parameters chosen for the drive pulse (left) and propagation medium (right). SG refers to Super-Gaussian order.

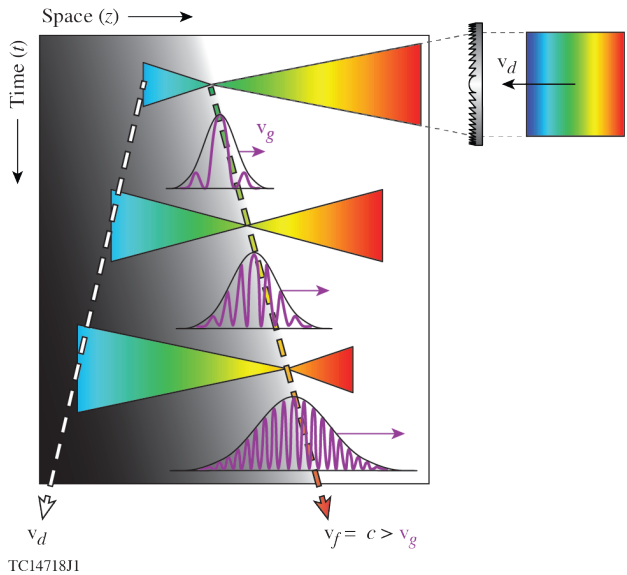

Figure 1. A schematic demonstrating photon acceleration in the ionization front of a flying focus. A negatively chirped drive pulse propagating at its group velocity, $v_d < 0$, forms a focus that counter-propagates at the velocity, $v_f = c$, triggering an ionization front traveling at $c$. The resulting electron density is indicated by shading. A witness pulse (drawn in purple) copropagates with the ionization front at velocity $v_g$ and continually upshifts in frequency.

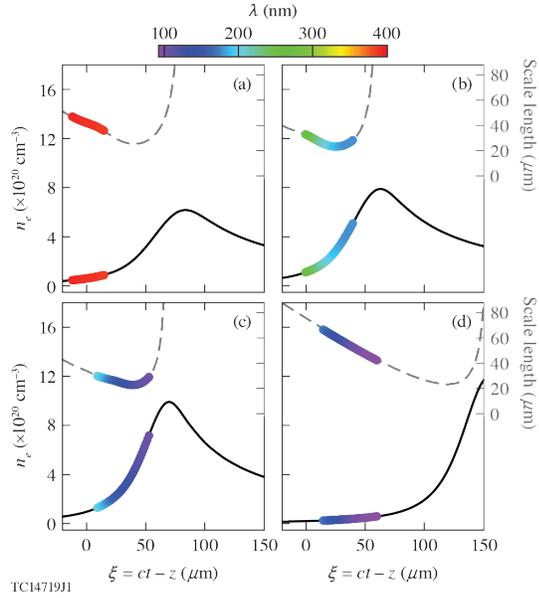

Figure 2. A series of snapshots of an 87 fs witness pulse with an initial wavelength $\lambda = 400$ nm copropagating with a temporal gradient in the electron density. The snapshots are taken at propagation distances of (a) 0.10, (b) 0.30, (c) 0.85, and (d) 1.05 cm and plotted in the moving frame, $\xi$. The pulse is modeled by photons initially spaced evenly in time over 87 fs. Each photon is represented by a circle colored to correspond to its vacuum wavelength (colorbar). The electron density and scale length, $L = n_e / \partial_\xi n_e$, are shown as solid black and dashed gray lines respectively.

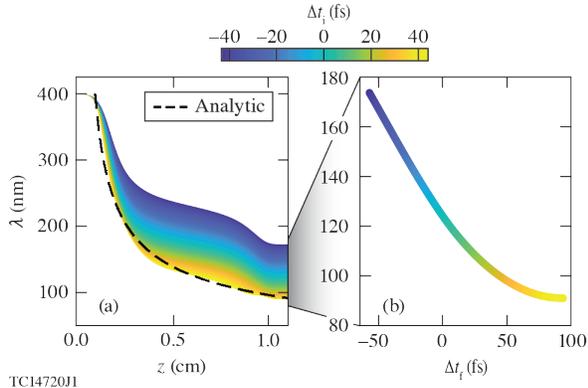

Figure 3. (a) A plot of vacuum wavelength, $\lambda$, over propagation distance, $z$, for a pulse modeled by a series of photons evenly distributed over an input duration of 87 fs, each with a color corresponding to the initial temporal location within the pulse, $\Delta t_i$ (colorbar). For comparison, the analytic case is plotted for the highest-frequency photon (dashed black line). (b) The distribution of wavelength over the final duration of the witness pulse, $\Delta t_f$, for the simulated case shown in (a).

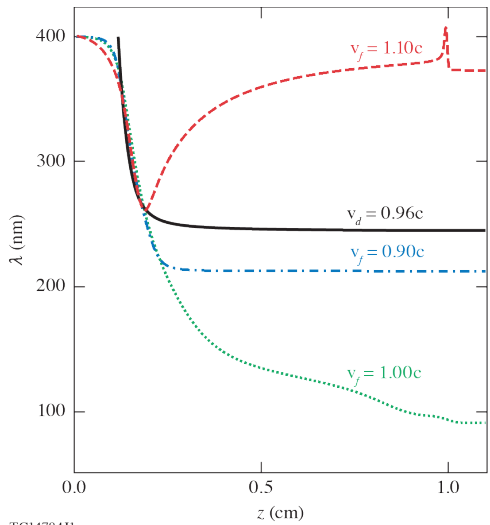

Figure 4. A plot of minimum vacuum wavelength over propagation distance, displayed for four separate ionization front velocities. Drawn in a green dotted line is a reproduction of the simulated case in (a), with a flying focus focal velocity of $v_f = c$; in a red dashed line is a super-luminal case in which $v_f = 1.1c$; in a blue dash-dotted line is a subluminal case in which $v_f = 0.9c$; in a black solid line is a case in which no flying focus is formed, and instead, the ionization front travels at the group velocity of the drive pulse $v_d = 0.96c$.